\documentclass[amssymb,twocolumn,aps]{revtex4}
\usepackage{graphics,color,graphicx,amsmath}
\usepackage{subcaption}
\usepackage{natbib}
\usepackage{verbatim}
\usepackage{graphicx}
\usepackage[above,below]{placeins}
\usepackage{float}
\usepackage{tabularx}
\usepackage{times}
\usepackage{blindtext}
\usepackage{url}
\usepackage{rotating}
\usepackage{mathtools}
\usepackage{threeparttable}

\begin{document}

\title{Advancing clustering methods in physics education research: A case for mixture models}

\author{Minghui Wang$^{1*}$~\footnote[0]{*Co-first authors}, Meagan Sundstrom$^{2*}$~\footnote[2]{Corresponding author: sundstrommeagan@gmail.com}, Karen Nylund-Gibson$^{3}$, and Marsha Ing$^{4}$}
\affiliation{
$^{1}$Department of Communication, University of California, Santa Barbara, Santa Barbara, CA
93106, USA\\
$^2$Department of Physics, Drexel University, Philadelphia, PA 19104, USA\\
$^{3}$Department of Education, University of California, Santa Barbara, Santa Barbara, CA
93106, USA\\
$^4$School of Education, University of California, Riverside, Riverside, CA 92521, USA}

\date{\today}

\begin{abstract}
Clustering methods are often used in physics education research (PER) to identify subgroups of individuals  within a population who share similar response patterns or characteristics. \textit{K}-means (or \textit{k}-modes, for categorical data) is one of the most commonly used clustering methods in PER. This algorithm, however, is not model-based: it relies on algorithmic partitioning and assigns individuals to subgroups with definite membership. Researchers must also conduct post-hoc analyses to relate subgroup membership to other variables. Mixture models offer a model-based alternative that accounts for classification errors and allows researchers to directly integrate subgroup membership into a broader latent variable framework. In this paper, we outline the theoretical similarities and differences between \textit{k}-modes clustering and latent class analysis (one type of mixture model for categorical data). We also present parallel analyses using each method to address the same research questions in order to demonstrate these similarities and differences. We provide the data and R code to replicate the worked example presented in the paper for researchers interested in using mixture models.
\end{abstract}

\maketitle

\section{Introduction}

Clustering methods are quantitative approaches to identifying distinct subgroups within a population that have become a common analysis technique in social science research~\cite{gutierrez2008research,harper2008they} and in physics education research (PER) specifically (e.g., Refs.~\cite{pond2017learning,battaglia2019unsupervised,quinn2020roles,wan2021TA,sundstrom2022intergroup,kortemeyer2024cheat}). These methods allow researchers to take a “person-centered approach” to data analysis, where the goal is to group together similar students (or other units of analysis) based on their multivariate response patterns (e.g., to a survey). This approach is in contrast to a “variable-centered approach,” where researchers examine relationships between variables across an entire dataset.

Within PER, a common method for identifying subgroups in a dataset is \emph{k}-means (or \emph{k}-modes, for categorical data), an algorithm that identifies clusters by iteratively defining cluster centers and minimizing the distance between individual observations and their closest cluster center (see Section~\ref{kmeanstheory} for further details)~\cite{quinn2020roles,sundstrom2022intergroup,kortemeyer2024cheat}. One study using \emph{k}-means clustering, for example, examined physics students’ use of online resources when completing their physics course assignments~\cite{kortemeyer2024cheat}. The researchers identified four types of users: those who mostly use peer discussions and help rooms, those who use online resources on homework but not on exams, those who heavily use the Internet for both homework and exams, and those who use all resources (including Artificial Intelligence) on all assignments. Another study using \emph{k}-means clustering investigated student roles when completing physics laboratory groupwork based on video observations of student behaviors~\cite{quinn2020roles}. The authors identified five subgroups of students. Each subgroup was defined by one behavior that students engaged in more than average: handling the equipment, using a laptop or personal device, using the desktop computer at the lab bench, writing on paper or in a notebook, and other behaviors (e.g., writing on whiteboards, off-task behaviors). The use of clustering methods in both of these PER studies provided a nuanced understanding of student behaviors.

Researchers are often also interested in how a student’s cluster membership relates to other variables, such as demographic information and performance. The \emph{k}-means algorithm definitively assigns each student to one cluster. Therefore, researchers typically take a post-hoc approach to this type of analysis. They first identify the clusters and assign each student to one cluster, and then use ANOVA, regression, or descriptive statistics to compare other variables across the clusters. In the study of physics students’ use of resources mentioned above, for instance, the authors compared the average homework and exam scores of students assigned to each type of user~\cite{kortemeyer2024cheat}. In the study of student roles in labs, the researchers used a chi-squared test of frequencies to compare the number of students of each gender identity that were assigned to each role~\cite{quinn2020roles}.

Despite its wide use, there are limitations to the \emph{k}-means clustering method. For example, the identification of clusters is procedural: the algorithm relies on a set of rules (i.e., minimizing distance between observations and their cluster center) that cannot be adapted. In addition, definitive cluster assignment means that \emph{k}-means clustering does not account for misclassification errors (i.e., students being assigned to the wrong cluster) when relating cluster membership to other variables (e.g., in a regression).

An alternative approach to clustering is mixture models, which are model-based. Under this method, the algorithm that creates the subgroups is flexible; for example, researchers can impose constraints on different model parameters. Mixture models also account for misclassification errors by assigning students probabilities of belonging to each identified subgroup (rather than definitive assignments). Researchers can directly integrate the resulting subgroups into a broader latent variable framework, where the identification of the subgroups (and the probabilities of each student belonging to each subgroup) and the relationships between students’ subgroup membership and any auxiliary variables are estimated simultaneously. 

While there is limited use of mixture models in PER thus far (see Ref.~\cite{commeford2022characterizing} for one example), there is great potential for leveraging these methods in the field as demonstrated by the studies that have already applied clustering methods~\cite{pond2017learning,battaglia2019unsupervised,quinn2020roles,wan2021TA,sundstrom2022intergroup,kortemeyer2024cheat}. This article, therefore, introduces one particular mixture modeling method---latent class analysis (LCA)---to the PER community with a worked example (including corresponding data and analysis scripts, see Ref.~\cite{github}). 

In the next section (Theory), we describe in further detail the theoretical similarities and differences between \emph{k}-modes clustering and LCA (because our worked example uses categorical data). In the following section (Application), we apply both methods to a data set related to undergraduate physics students’ social support networks to demonstrate these similarities and differences. We conclude with a summary of the two methods and recommendations for researchers.

\section{Theory}

Cluster analyses assume that the population being studied is not homogeneous, but rather consists of subgroups with different attitudes, traits, predispositions, or behaviors. A wide range of statistical methods have been used to identify these subgroups, including \emph{k}-means (or \emph{k}-modes) clustering~\cite{macqueen1967some}, hierarchical clustering~\cite{murtagh2012algorithms}, Gaussian mixture models~\cite{reynolds2009gaussian}, Density-Based Spatial Clustering of Applications with Noise (DBSCAN)~\cite{schubert2017dbscan}, deep embedding clustering~\cite{xie2016unsupervised}, and latent profile analysis (or latent class analysis)~\cite{muthen2001latent,masyn2013latent-key}. In this paper, we focus on \emph{k}-means/\emph{k}-modes clustering, given its prevalence in PER, and latent class analysis, the method we aim to illustrate. In the following subsections, we describe the algorithm underlying each method's identification of subgroups and then compare and contrast the two methods more broadly.

\subsection{\emph{K}-means and \emph{k}-modes clustering}
\label{kmeanstheory}

The \emph{k}-means algorithm partitions a set of observations into a pre-specified number of subgroups, called clusters, based on a set of continuous input variables~\cite{macqueen1967some}. The algorithm works as follows: first, based on the decided number of clusters (\textit{k}), the algorithm assigns each observation a random cluster number from 1 to \textit{k}, which serves as the starting cluster assignment (Step 1). Next, the algorithm calculates the ``centroid" (or center) of each cluster, computed as the vector of the variable means for all observations assigned to that cluster (Step 2). Each observation is then reassigned to a new cluster based on the closest centroid, where closeness is typically defined using Euclidean distance (Step 3)~\cite{james2013introduction}.  This algorithm can be described as minimizing the sum of the square of the distances from each point to its respective cluster’s centroid:
\begin{equation*}
    \underset{C_1,...C_k}{\text{minimize}}
     \biggl\{ 
    \sum_{k=1}^{K} 
    \sum_{x_i \in C_k}
    ||x_{i} - \mu_{k}||^2
    \biggl\}
\end{equation*}
where $C_k$ denotes the $k^{\text{th}}$ cluster, $K$ denotes the total number of clusters, $x_i \in C_k$ denotes observations assigned to cluster \emph{k}, $\mu_k$ denotes the centroid of cluster \emph{k}, and $|| \bullet ||^2$ denotes squared Euclidean distance.

The goal of \emph{k}-means clustering is to minimize within-cluster variation, such that each observation is more similar to observations in the same cluster than to other observations in different clusters. The algorithm iterates Steps 2 and 3 until the cluster assignments no longer change. Mathematically, each iteration is guaranteed to decrease the within-cluster variation.

While the \emph{k}-means clustering algorithm~\cite{macqueen1967some} is designed for continuous data, Huang~\cite{huang1997fast} extended the algorithm to handle categorical data. This \emph{k}-modes algorithm also iteratively defines centroids of clusters, assigns observations to clusters, and updates the centroids until the cluster assignments no longer change. However, in \emph{k}-modes the cluster centroids are defined by the mode, or the most common value, of each variable for observations assigned to the cluster. Instead of Euclidean distance, simple-matching distance is used to determine the similarity of two observations. Simple-matching distance is computed by counting the number of variables for which two observations have different values~\cite{huang1997fast}.

\subsection{Latent class analysis}

Latent class analysis (LCA) is a model-based method used to identify unobserved latent subgroups based on categorical data, where each subgroup, or “class,” is characterized by a unique pattern of response probabilities to the input variables, or ``indicators." The method is part of a larger modeling framework referred to as mixture modeling~\cite{muthen2001latent,masyn2013latent-key}, where the latent class variable (which creates the subgroups) is a categorical latent variable and its relation to other auxiliary variables, including latent variables (e.g., factors as in factor analysis), can be estimated. To determine the subgroups, the LCA algorithm estimates two sets of parameters: 
\begin{enumerate}
    \item Structural parameters: the proportion of the population belonging to each class, $P(c = k)$, which indicates the relative class size, and
    \item Measurement parameters: the conditional item probabilities, or the probability that a student in class \textit{k} would endorse a specific indicator \textit{j}, $P(u_j = 1|c = k)$~\cite{nylund2018ten}.
\end{enumerate}
The basic LCA model assumes conditional independence, meaning that the latent class variable creating the subgroups explains all of the shared variance in the observed indicators. 

To estimate the model, the LCA algorithm iterates between expectation (E-step), where it computes the probability that each individual observation belongs to each latent class, and maximization (M-step), where it updates the model parameters $P(c = k)$ and $P(u_j = 1|c = k)$ to improve the fit. In more detail, the algorithm can be described as follows: as a starting point (initialization), observations are split into \textit{k} classes at random, where the \textit{k}-classes are all roughly equal in size. In the Expectation step (E-step), the algorithm estimates the probability that each individual belongs to each latent class with the current parameter estimates. This is done by computing “posterior” probabilities of class assignment given the current parameter estimates. Next, in the Maximization step (M-step), the algorithm maximizes the expected log-likelihood function by updating the parameter estimates based on the expected values obtained in the E-step. The log-likelihood function optimized by the EM algorithm is expressed as:
\begin{equation*}
    \text{log} L (\theta) = 
    \sum_{i=1}^N 
    \text{log}
    \biggl(
    \sum_{k=1}^K P(c=k)
    \prod_{j=1}^{J} P(u_{ij} = 1|c = k) 
    \biggl)
\end{equation*}
where $N$ is the number of observations, $K$ is the number of classes, and $J$ is the number of indicators. The E- and M-steps are repeated iteratively until the log-likelihood function is stable and converges on a solution (each software has a threshold at which it determines that the change in the likelihood is small enough). 

\subsection{Comparing \emph{k}-modes clustering and LCA}

The input variables in our worked example (presented next) are all binary, therefore we focus our comparison of clustering methods on \emph{k}-modes clustering and LCA. However, all of the discussion points also hold true for the analogous clustering methods for continuous indicators: \emph{k}-means clustering and latent profile analysis.

There are several similarities between \emph{k}-modes clustering and LCA (Table~\ref{comparemethods}). First, both methods assume that the population is heterogeneous and that identifying meaningful subgroups (referred to as clusters in \emph{k}-modes and classes in LCA) better captures meaningful heterogeneity than treating the whole population as one homogeneous group. Second, the primary goal of both methods is to minimize within-group differences: observations assigned to the same subgroup should be as similar as possible to each other and observations in different subgroups should be as distinguishable from each other as possible. Additionally, both methods partition the data by assuming that each data point belongs to exactly one cluster (\emph{k}-modes) or class (LCA), which implies that clusters or classes are mutually exclusive and exhaustive. Both approaches also use iterative algorithms to arrive at the optimal groups: \emph{k}-modes clustering iteratively updates centroids until convergence, and LCA iteratively estimates parameters using the Expectation-Maximization (EM) algorithm. The implication of the iterative process is that both methods are susceptible to converging at local, rather than global, solutions. As such, both approaches require multiple runs with different initializations to ensure that the model estimates are based on the best solution.

 Despite these similarities, \emph{k}-modes clustering and LCA differ in fundamental ways (Table 1). The key distinction is that LCA is a model-based clustering technique grounded in probability theory. In other words, LCA assumes that “the data are generated by a mixture of underlying probability distributions”~\cite[p. 90]{vermunt2002latent}. This model-based approach has several advantages. First, when identifying the subgroups, LCA is less sensitive to outliers than \textit{k}-modes clustering. \emph{K}-modes clustering is an outlier-intolerant algorithm--an outlier can cause the algorithm to create an additional (but uninformative) cluster or pull a cluster centroid towards it, distorting the cluster boundaries~\cite{gan2017k,chawla2013k}. In contrast, the model-based LCA approach estimates the conditional probability that an observation belongs to each latent class, making it more robust to outliers. 

 Second, researchers using LCA have the flexibility to manually impose restrictions on model parameters to reflect their theoretical assumptions. For example, researchers can test measurement invariance (i.e., whether the latent classes have consistent meaning across conditions) by constraining parameters across groups and comparing the fits of nested models. Researchers may also conduct confirmatory LCA, where they explicitly specify the expected class structure based on prior theory. Such imposed constraints are not possible in \emph{k}-modes clustering. 

  Next, LCA has several associated model fit indices that help researchers choose the optimal number of classes (i.e., several information criteria directly compare different models to one another, described further in the Application section) and several classification diagnostics that evaluate the overall goodness-of-fit of the models to the data. \emph{K}-modes clustering, on the other hand, is a heuristic method: there are a few validation indices for determining the optimal number of clusters (described further in the Application section) and no classification diagnostics. 
 
 Finally, LCA is particularly useful when the classes are considered in a larger variable system and relations among the identified classes and other variables are hypothesized. When relating the classes to auxiliary variables (similar to how variables are conceptualized in a broader modeling approach, such as structural equation modeling), classification errors are accounted for~\cite{nylund2014latent}. That is, an individual’s contribution to the mean of an auxiliary variable of class \textit{k} is weighted by their posterior probability of belonging to that class. In contrast, \emph{k}-modes clustering assigns observations to clusters with weights of either 0 or 1, which does not account for misclassification errors. Information about potential misclassification of individuals into clusters is not carried forward in later analyses (e.g., an ANOVA comparing the means of an auxiliary variable across clusters), which may introduce bias in these analyses.

\begin{table*}[t]
  \caption{Comparison between \textit{k}-modes clustering and LCA.\label{comparemethods}}
  \begin{ruledtabular}
    \begin{tabular}{p{0.3\textwidth}p{0.25\textwidth}p{0.25\textwidth}}
  & \textit{K}-modes clustering & LCA \\ 
 \hline
 Assumes heterogeneity & Yes & Yes \\ \\
 Statistical foundation & Minimizes within-cluster variation & Maximizes likelihood \\ \\
  Susceptible to converging on local (rather than global) solution   & Yes & Yes \\ \\
 Model-based & No & Yes\\ \\
 Sensitivity to outliers & High & Low\\ \\
 Manual constraints allowed & No & Yes\\ \\

 Model fit indices & Yes, but limited & Yes\\ \\
  Model classification diagnostics & No & Yes\\ \\
 Cluster/class assignment type&
Hard (the weight is either 0 or 1) &
Soft (posterior probability) \\ \\

Auxiliary variable integration & Not supported natively (typically done post-hoc) & Supported in the larger latent variable framework \\ \\
Computational load & Computationally efficient  & Computationally laborious \\ \\
Tools &
R: \textit{klaR}; 
Python: \textit{kmodes} & Mplus; Latent Gold;
R: \textit{poLCA}; Python: \textit{StepMix}, \textit{scikit-learn}\\

    \end{tabular}
  \end{ruledtabular}
\end{table*}
 
There are also challenges when applying LCA as compared to \emph{k}-modes clustering. The accessibility of software to run LCA is limited. Both Mplus and Latent Gold are proprietary and require purchase, potentially limiting access for researchers with budget constraints. Free mixture modeling software in R, such as the \textit{poLCA} package, can be useful, but such packages are limited relative to the modeling capabilities available in Mplus or Latent Gold. In contrast, software to run \emph{k}-modes clustering is widely accessible through free software such as the \textit{klaR} package in R and \textit{kmodes} package in Python. LCA is also more computationally laborious compared to \emph{k}-modes. \emph{K}-modes clustering is often more practical for large sample sizes due to its efficiency, scalability, and lower resource demands. Lastly, mixture modeling necessitates a more advanced understanding of statistical concepts and methods. For example, LCA benefits from multiple sources of fit information that require an understanding of how to triangulate the information of these fit indices to make decisions about the optimal number of classes. 

In terms of the clusters or classes that each method identifies, several studies in social science have directly compared the subgroups that result from applying \emph{k}-means/\emph{k}-modes clustering and LCA to the same data set. Grant and colleagues~\cite{grant2020use}, for example, used LCA and \emph{k}-means methods on medical record data to identify distinct subgroups of medically complex patients (they transformed the categorical data used in LCA to continuous data for \emph{k}-means clustering). Their LCA resulted in a seven-class solution, while \emph{k}-means clustering yielded an eight-cluster solution; however, both analyses resulted in qualitatively similar groupings. In another study, Papachristou and colleagues~\cite{papachristou2018congruence} applied both LCA and \emph{k}-modes clustering to identify subgroups of oncology patients based on the symptoms they self-reported on a survey. They identified a similar set of subgroups using both methods, but the classes from the LCA were better separated from one another (i.e., more distinct) than the clusters from \emph{k}-modes. Similarly, Magidson and Vermunt~\cite{magidson2002latent} compared the performance of LCA and \emph{k}-means using simulated data with known true subgroup membership. Results indicated that LCA substantially outperformed the \emph{k}-means algorithm in assigning observations to the correct subgroup. These findings point to possible advantages of LCA over \emph{k}-modes clustering; however, we run both clustering methods in our worked example to clearly illustrate their differences and directly compare their subgroup identification in the context of physics education.

\section{Application}

In this section, we describe the process and results of applying both \emph{k}-modes clustering and LCA to a physics education data set. We perform each analysis on its own using the current best practices for each method, as a researcher would if they were to select one or the other for their study. Our goal is to demonstrate the similarities and differences between the two methods, described in the previous section, in the context of research questions relevant to the PER community. De-identified data and analysis scripts for both analyses are available at Ref.~\cite{github}.

\subsection{Background and research questions}

We examine the sources of social support that women and other gender minorities draw upon during their undergraduate physics program. Women and other gender minorities are significantly under-represented in physics courses and majors~\cite{clark2005,sax2016,porter2019women}, and the factors that are related to these students’ success in the discipline---including social support---are of increasing interest to the PER community~\cite{franklinreasons,gutzwa2024nets,hatcher2025egocentricmixedmethodssnaanalyzing,franklinpersist}.

Our analysis is motivated by Tinto’s Interaction Model of Student Retention~\cite{tinto1975dropout}, which contends that undergraduate students are more likely to persist (e.g., complete their degree program) if they are socially integrated in their institution, including with other students, faculty, staff, and extracurricular groups. Other studies suggest that additional sources of social support outside of the institution, such as family members, are also important for student success~\cite{wicker2023web}. Each of these groups of people may provide students with different types of resources and support (e.g., emotional, informational, instrumental) that contribute to their success. Faculty members, for example, may provide informational support by telling students about possible career options in their discipline, while family members may provide emotional support by showing care for students during challenging moments.

Prior research has started to investigate the role of social support networks in the academic trajectories of historically under-represented college students, including women and other gender minorities in physics. One body of work identifies that historically under-represented students who have social networks spanning different types of support tend to feel a stronger sense of belonging and a stronger discipline-specific identity, and are more likely to complete their degree, than students who rely on only one or a few types of social support~\cite{wicker2023web,gutzwa2024nets,kim2018science}. These primarily qualitative studies (e.g., interviewing relevant populations about their sources of social support and how they contributed, or not, to their academic success) provide important narratives about the role of social support networks, but the results are not intended to be generalizable. Other work has quantitatively examined the extent to which different sources of social support relate to student outcomes (e.g., degree completion); however, this body of literature examines correlations between each individual source of support and student outcomes (e.g., using multiple regression models in a variable-centered approach)~\cite{mishra2020social}. 

To our knowledge, no quantitative studies have taken a person-centered, multivariate approach to identifying the various \textit{combinations} of social support students draw on and how these combinations relate to student outcomes. Yet the qualitative studies mentioned above point to the existence and importance of these combinations: some students may receive support from peers and faculty in their department \textit{and} from friends, family, and others outside of their institutions, while other students may have support only from peers and faculty in their department or only from friends and family. In physics specifically, students of different gender identities may draw on different combinations of social support and these combinations are likely predictive of their physics identity, or the extent to which they feel like they are a “physics person”~\cite{gutzwa2024nets,hazari2010connecting}. Physics identity has been shown to positively correlate with student persistence in undergraduate physics courses and physics careers~\cite{hazari2010connecting}. Identifying the most effective combinations of social support and who may be lacking them, therefore, may inform the design of targeted instructional interventions (e.g., aimed at helping students find and maintain certain types of social support) that have the potential to improve gender representation in physics.

Our worked example of \emph{k}-modes clustering and LCA is guided by the following research questions:
\begin{enumerate}
    \item What combinations of social support do undergraduate women and gender minorities in physics draw upon?
    \item  How does students’ combination of social support relate to their gender identity and physics identity?
\end{enumerate}
We note that we do not aim to completely address these research questions; rather, we use them to illustrate a practical application of clustering methods in PER. As such, we simplify our analysis to only include one gender identity as a predictor of cluster/class membership---whether or not students identify as non-binary---and one outcome variable predicted by cluster/class membership---physics identity.


\subsection{Data collection and preparation}

We collected survey data from participants at the 2025 Conference for Undergraduate Women and Gender Minorities in Physics (CU*iP). CU*iP is an annual, three-day conference run by the American Physical Society. The conference is structured as several concurrent, regional conferences, such that students attend a site close to their home institution. About 50 to 200 students attend each site (there are typically about 10 sites), with a total attendance of around 500 to 2000 participants per year. The purpose of the conference is for women and gender minorities pursuing physics degrees to network with physics peers from their geographic area, to present their research in a supportive environment, and to learn about different career paths in physics. The cost to participate is low (\$45 registration fee) and often supported by the American Physical Society and/or participants’ home institution. The majority of women and gender minorities in physics attend at least one CU*iP during their undergraduate program. As we collected our data from one conference year, our data likely do not capture the experiences of \textit{all} women and gender minorities in physics. However, the data likely capture much of the \textit{variation} in experiences of this student population because the participants come from many different institutions and geographic regions of the United States.

The data analyzed here come from the conference registration survey, which is given to participants when they register to attend the conference a few weeks before the event. To identify combinations of social support, we asked the following question on the registration survey: “What groups and/or people have you drawn support from in your journey through undergraduate physics (select all that apply)?” Participants could select from the following options, which we curated based on prior qualitative studies of social support for women and other gender minority physicists~\cite{gutzwa2024nets}:
\begin{itemize}
    \itemsep -0.1cm
    \item Family members
    \item Physics classmates/peers
\item Friends/peers not in physics
\item High school physics teacher(s)
\item Faculty (not physics) at your institution
\item Physics faculty at your institution
\item Research group/research mentors
\item Women in physics or STEM groups
\item Gender inclusive physics/STEM groups
\item Religious/faith-based organizations
\item Other identity-based affinity groups
\item Social communities (e.g., book clubs, fitness group, game club)
\item Mental health professionals
\item Other, please specify (provides an open text box)
\end{itemize}

We also asked students to self-report their gender identity using the following prompt: “How would you describe your gender identity (select all that apply)?” Students could select from the following options:
\begin{itemize}
    \itemsep -0.1cm
    \item Woman
\item Man
\item Genderqueer
\item Agender
\item Transgender
\item Cisgender
\item A gender not listed (provides an open text box)
\item Prefer not to disclose
\end{itemize}

We measured physics identity through two survey prompts developed in prior work~\cite{hazari2020context}: “With respect to a physics community, to what extent do you see yourself as a physicist?” and “To what extent do you see yourself as an exemplary physics student?” Each prompt was a 5-point Likert-scale question with options ranging from 0 (“not at all”) to 4 (“very much so”). Respondents could only select one option for each of these two prompts.

\begin{figure}[b]
    \centering
    \includegraphics[width=3.4in,trim={0 0 0 0 }]{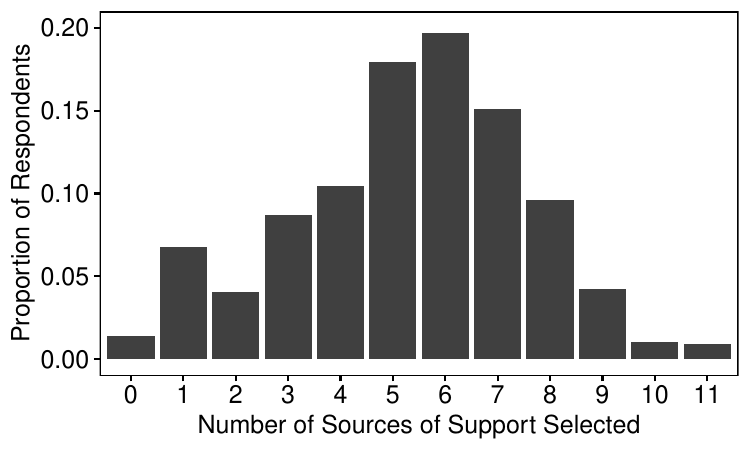}
    \caption{Distribution of the number of sources of support selected by each respondent on the survey (\textit{n} = 567).}
    \label{fig:Hist}
\end{figure}

\begin{figure}[b]
    \centering
    \includegraphics[width=3.4in,trim={0 0 0 0 }]{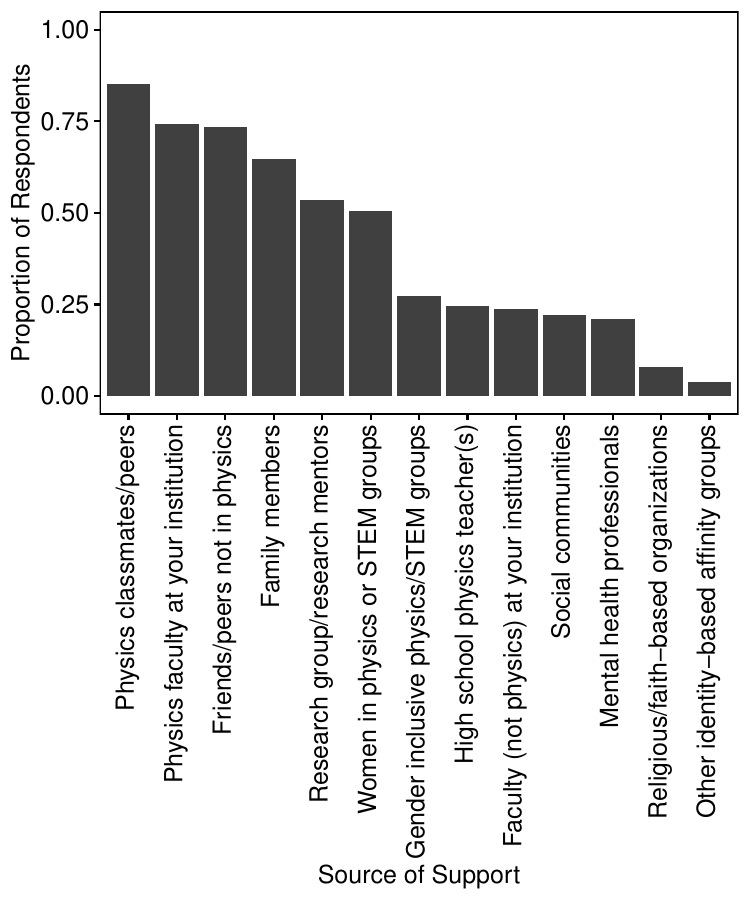}
    \caption{Proportion of respondents who selected each source of support on the survey (\textit{n} = 567). Respondents could select as many sources of support as they wanted.}
    \label{fig:Bars}
\end{figure}

A total of 599 participants filled out the registration survey. Participants were excluded from our analysis if at least one of the following was true: they did not respond to the gender identity prompt, they only selected “prefer not to disclose” on the gender identity prompt, they only selected “man” on the gender identity prompt, they only selected “man” and “cisgender” on the gender identity prompt, and they did not respond to at least one of the two physics identity prompts. 567 participants (96\% of all respondents) were included in our analysis. 

We processed responses to the social support prompt by turning the first 13 options into 13 separate, binary variables: 1 if the participant selected that option and 0 if the participant did not select that option. We excluded the final option (“Other, please specify”) from our analysis because it involved a free response (i.e., complex and nuanced ideas) and only eight respondents selected this option. 98.6\% of respondents selected at least one source of social support and respondents selected, on average, 5.3 (standard deviation, SD = 2.3) sources of social support (Fig.~\ref{fig:Hist}). Physics peers were the most commonly identified source of social support, followed by physics faculty, non-physics peers, and family members (Fig.~\ref{fig:Bars}). Religious/faith-based organizations and other identity-based affinity groups were the least common sources of social support reported by the respondents.

We created three separate, non-exclusive variables for gender identity: “Woman” if the participant selected “Woman,” “Transgender” if the participant selected “Transgender,” and “Non-binary” if the participant selected “Genderqueer,” “Agender,” and/or “A gender not listed.” 498 (88.5\%) participants identified as women, 38 (6.7\%) identified as transgender, and 93 (16.5\%) identified as non-binary. 

We calculated each participant’s physics identity as the average of their Likert responses to the two physics identity prompts. Participants reported, on average, a physics identity of 2.5 (SD = 0.84).

\subsection{\textit{K}-modes clustering}


\subsubsection{Cluster identification}

We performed the \emph{k}-modes clustering using the \textit{klaR} package in R (Version 4.3.2)~\cite{huang1997fast}. The clustering was performed iteratively for a range of cluster values (\textit{k}) from 1 to 10. The maximum number of iterations allowed was set as 300 and a random seed was set for reproducibility.

\begin{figure}[b]
    \centering
    \includegraphics[width=3.4in,trim={0 0 0 0 }]{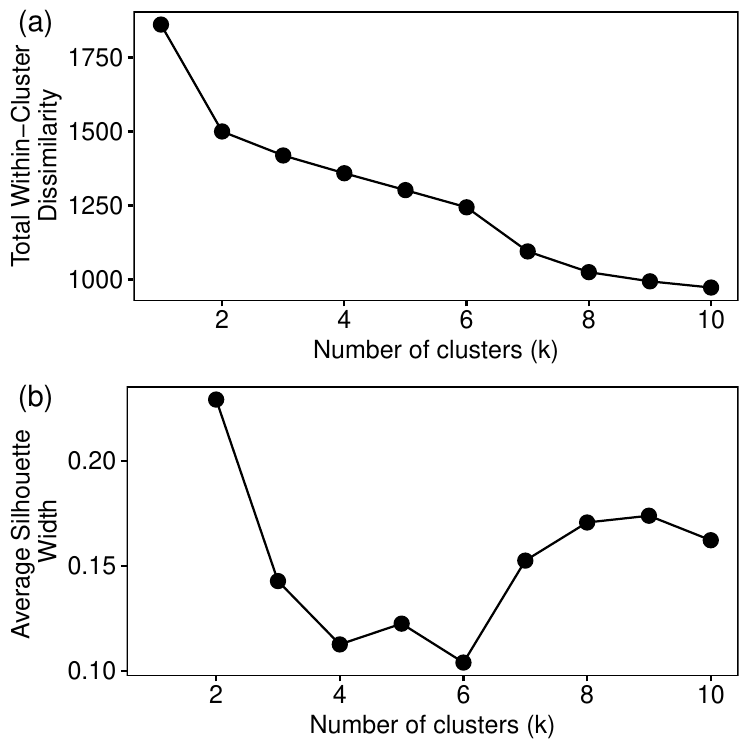}
    \caption{Elbow plots of (a) total within-cluster dissimilarity and (b) average silhouette width for \emph{k}-modes clustering.}
    \label{fig:Elbows}
\end{figure}

We used two metrics to determine the optimal number of clusters: total within-cluster dissimilarity and average silhouette width (Fig.~\ref{fig:Elbows})~\cite{rousseeuw1987silhouettes}. Total within-cluster dissimilarity measures the number of variables for which each observation has a different value than their respective cluster's mode, calculated using simple matching distance:
\begin{equation*}
    \sum_{k=1}^K \sum_{x_i \in C_k} 
    \sum_{j=1}^m
    \delta(x_{ij},\text{mode}(C_{k}))
\end{equation*}
where $K$ is the total number of clusters, $m$ is the number of input variables, and $ \delta(x_{ij},\text{mode}(C_k))$ is 0 if $x_{ij} = \text{mode}(C_k)$ and 1 otherwise. For this metric, lower values indicate more cohesive clusters and, therefore, indicate superior solutions. The elbow point on the plot of values of the metric for each number of clusters, where increasing the number of clusters \textit{k} does not significantly reduce the within-cluster dissimilarity, suggests the optimal number of clusters. For our dataset, the elbow point occurs at two clusters (Fig.~\ref{fig:Elbows}a). Notably, the shape of the elbow plot was sensitive to the chosen random seed. Elbow plots generated with different random seeds suggested that the optimal solution varied, with the two-cluster solution as the elbow point in some cases and the three-cluster solution as the elbow point in others. 

The average silhouette width compares intra-cluster and inter-cluster dissimilarity. For each observation, the intra-cluster dissimilarity ($a(i)$) is its average simple matching distance to all other observations in the same cluster, while the inter-cluster dissimilarity ($b(i)$) is its average simple matching distance to all observations in the nearest neighboring cluster. The silhouette score for each observation is then calculated as 
\begin{equation*}
S(i)= [b(i)-a(i)]/[\text{max}(a(i), b(i)], 
\end{equation*}
and the average silhouette width is obtained by averaging $S(i)$ across all observations in the dataset. Values close to 1 (--1) indicate strongly (weakly) separated clusters. For our dataset, the average silhouette width is 0.23 for the two-cluster solution and 0.14 for the three-cluster solution (Fig.~\ref{fig:Elbows}b). Therefore, both metrics point to the two-cluster solution as the most appropriate.

\begin{figure}[b]
    \centering
    \includegraphics[width=3.4in,trim={0 0 0 0 }]{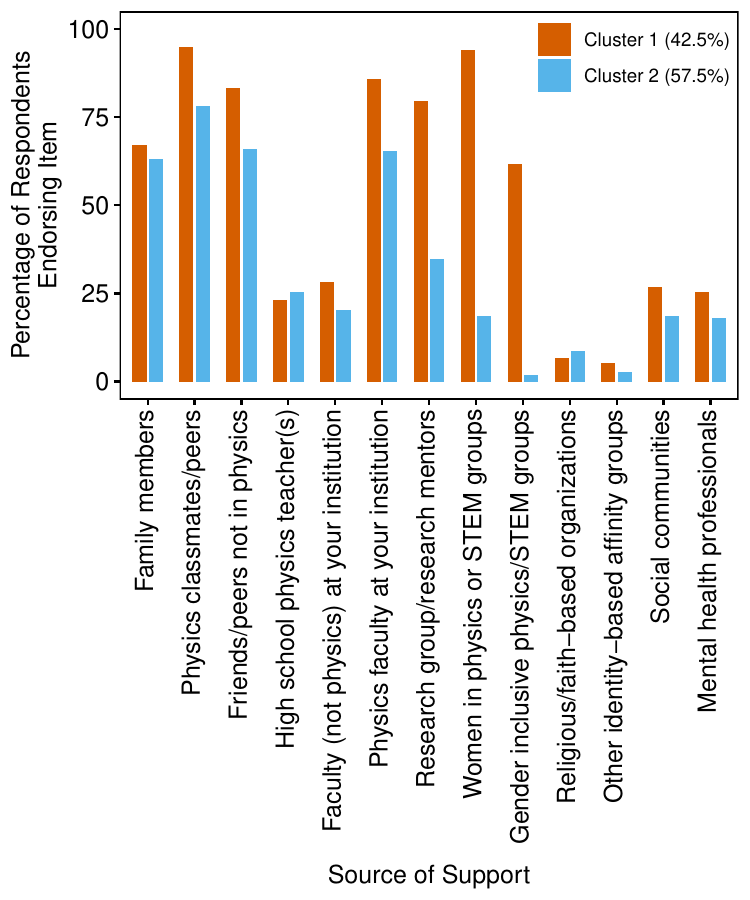}
    \caption{Two-cluster solution for \emph{k}-modes clustering. Percentages in the legend indicate the percent of students assigned to each cluster (\textit{n} = 567).}
    \label{fig:2Cluster}
\end{figure}

We characterized each of the two resulting clusters based on the percentage of individuals in the cluster who selected each source of social support (Fig.~\ref{fig:2Cluster}). Cluster 1, which we label as “High professional and identity-based support,” consists of 42.5\% (\textit{n} = 241) of the sample. Students in this cluster have high item endorsement percentages across nearly all support sources, especially physics classmates or peers (95.0\%), women in physics or STEM groups (94.2\%), physics faculty at their institution (85.9\%), friends or peers not in physics (83.4\%), and research group/research mentors (79.7\%). This cluster, therefore, consists of students who receive social support from multiple different sources, including institutional, interpersonal, and identity-based sources. 

Cluster 2, which we label as “High professional and low identity-based support,” consists of 57.5\% (\textit{n} = 326) of the sample. Students in Cluster 2 show moderate-to-high endorsement across most support sources, including physics classmates or peers (78.2\%), friends or peers not in physics (66.0\%), and physics faculty at their institution (65.3\%). Notably, however, the percentage of endorsements for receiving support from women in physics or STEM groups (18.7\%) and from other gender inclusive groups (1.8\%) is much lower in Cluster 2 compared to Cluster 1. There is also a lower percentage of endorsements for receiving support from research group members and research mentors in Cluster 2 compared to Cluster 1.

\begin{table*}[t]
  \caption{Model fit summary for latent class analysis (see plot of values in Fig.~\ref{fig:IC} in the Appendix). Bold indicates the optimal value for each model fit index or likelihood ratio test (LRT).\label{modelfit}}
  \begin{ruledtabular}
    \begin{tabular}{lccccccccccc}
  
  &&&&&&&&&&& Smallest\\
   & & & & & \multicolumn{4}{c}{Model Fit Indices} & \multicolumn{2}{c}{LRTs} & Class\\
   \cline{6-9}
\cline{10-11} \cline{12-12}
 Classes  & npar & \textit{LL} & \% Converged & \% Replicated & BIC & aBIC & CAIC & AWE & VLMR & BLRT & n (\%)\\ 
 \hline
 1-Class & 13 & --3,840.13 & 100\% & 100\% & 7,762.68 & 7.721.41 & 7,775.68 & 7,884.11 & --- & --- & 567 (100\%)\\
 2-Class & 27 & --3,647.65 & 98\% & 98\% & 7,466.49 & 7,380.77 & \textbf{7,493.49} & \textbf{7,718.68} & $<$0.001 &$<$0.001 & 148 (26.1\%)\\
3-Class & 41 & --3,596.93 & 85\% & 95\% & \textbf{7,453.81} & \textbf{7,323.65} & 7,494.81 & 7,836.76 & \textbf{$<$0.001}&$<$0.001 & 120 (22.9\%)\\
4-Class & 55 & --3,576.71 & 23\% & 34\% & 7,502.15 & 7,327.55 & 7,557.15 & 8,015.87 & 0.68 & \textbf{0.04}& 101 (17.8\%)\\
5-Class & 69& --3,558.81 & 17\% & 10\% & 7,555.10 & 7,336.06 & 7,624.10 & 8,199.59 & 0.38 & 0.14 & 20 (3.5\%)\\
6-Class & 83 & --3,542.47 & 11\% & 19\% & 7,611.19 & 7,347.71 & 7,694.19 & 8,386.44 & 0.34 & 0.14 & 18 (3.2\%)\\
7-Class & 97 & --3,526.55 & 8\% & 2\% & 7,668.12 & 7,360.19 & 7,765.12 & 8,574.14 & 0.10 & 0.16 & 17 (3.0\%) \\
     
    \end{tabular}
  \end{ruledtabular}
  \begin{tablenotes}
      \small
      \item Note: npar = number of parameters; \textit{LL} = model log likelihood; \% Converged = percentage of runs where the model converged on a solution; \% Replicated = percentage of runs where the best log-likelihood was replicated; BIC = Bayesian information criterion; aBIC = sample size adjusted BIC; CAIC = consistent Akaike information criterion; AWE = approximate weight of evidence criterion; VLMR = Vuong-Lo-Mendell-Rubin adjusted likelihood ratio test \textit{p}-value; BLRT = bootstrapped likelihood ratio test \textit{p}-value; Smallest Class = number of observations in the smallest class.
    \end{tablenotes}
\end{table*}

\subsubsection{Relating social support cluster membership to gender identity and physics identity}

We related students’ cluster membership to other variables using a post-hoc analysis as in prior work~\cite{quinn2020roles,kortemeyer2024cheat}.  First, we used logistic regression to examine the relationship between student gender identity, particularly non-binary status, and \emph{k}-modes cluster membership (recall that to simplify our demonstration of these clustering methods, we only focus on the non-binary gender identity as a predictor of cluster membership). Cluster membership was set as the dependent variable and non-binary status was set as the independent variable. The results indicate that non-binary status is not a significant predictor of \emph{k}-modes cluster membership. That is, individuals identifying as non-binary show no difference in the likelihood of being assigned to Cluster 1 (High professional and identity-based support) versus Cluster 2 (High professional and low identity-based support; $B$ = --0.36, $p$ = 0.11).

We conducted an analysis of covariance (ANCOVA) to examine whether students' cluster membership predicts their physics identity, while controlling for non-binary status. The results indicate that cluster membership has a significant effect on physics identity; specifically, individuals in Cluster 1 (High professional and identity-based support) report stronger perceptions of their physics identity than individuals in Cluster 2 (High professional and low identity-based support; $B$ = --0.20, \textit{t}(560) = --2.79, $p$ $<$ 0.01). Non-binary status is not a significant predictor of physics identity in this model ($B$ = --0.14, \textit{t}(560) = --1.50, $p$ = 0.14).

\subsection{Latent class analysis}

\subsubsection{Class enumeration}

We estimated LCA models using maximum likelihood estimation with robust standard errors in MplusAutomation in R (Version 4.3.2)~\cite{hallquist2018mplusautomation}. We estimated the models with 200 random sets of starting values, as recommended by Hipp and Bauer~\cite{hipp2006local}, to ensure that the model converged on a global rather than a local solution. The algorithm first ran all 200 models for a fixed number of iterations and then selected the 100 models with the best log-likelihood values for full iteration. For models with four to seven classes, however, the convergence rate and best log-likelihood replication rate were both below 50\% (Table~\ref{200con} in the Appendix). To increase the possibility of reaching the global maximum, we increased the number of random starts to 2000 initial and 500 final iterations for these models. Although replication rates remained relatively low, the best log-likelihood values showed less variation, suggesting that the models likely converged on the right solution (Table~\ref{2000con} in the Appendix).

Based on current recommendations~\cite{masyn2013latent-key,nylund2018ten}, we considered four information criteria (IC) and two likelihood ratio tests (LRT) to choose the optimal number of classes. Specifically, we used the Bayesian information criterion (BIC), sample size adjusted BIC (aBIC), consistent Akaike information criterion (CAIC), and approximate weight of evidence criterion (AWE). For these four ICs, low values indicate superior model fit. The bootstrapped likelihood ratio test (BLRT) and the Vuong-Lo-Mendell-Rubin adjusted likelihood ratio test (VLMR-LRT) indicated if a specified \textit{k}-class model is significantly better than a \textit{k}-1 class model in terms of model fit. A significant \textit{p}-value indicates a significant improvement in the model fit of the \textit{k}-class model. A non-significant \textit{p}-value suggests that the \textit{k}-1 class model is preferred, as the two models have comparable model fit while the \textit{k}-1 class model is more parsimonious.

We triangulated these fit indices, and adhered to rules of parsimony, to select the optimal solution for our dataset, as no single fit index can solely determine the best model (Table~\ref{modelfit} and Fig.~\ref{fig:IC} in the Appendix). BIC and aBIC are minimized in the three-class model, while CAIC and AWE are minimized in the two-class model. However, the CAIC values for the two-class and three-class models are similar, suggesting only a minimal improvement in model fit of the two-class model in the three-class model. VLMR-LRT achieved a nonsignificant \textit{p}-value in a four-class model, which supports the three-class solution. BLRT achieved a nonsignificant \textit{p}-value in a five-class model, which supports the four-class solution. Multiple fit indices (BIC, aBIC, and VLMR-LRT) support the three-class model, and the CAIC does not strongly favor the two-class model, so we chose the three-class model.

\begin{figure}[b]
    \centering
    \includegraphics[width=3.4in,trim={0 0 0 0 }]{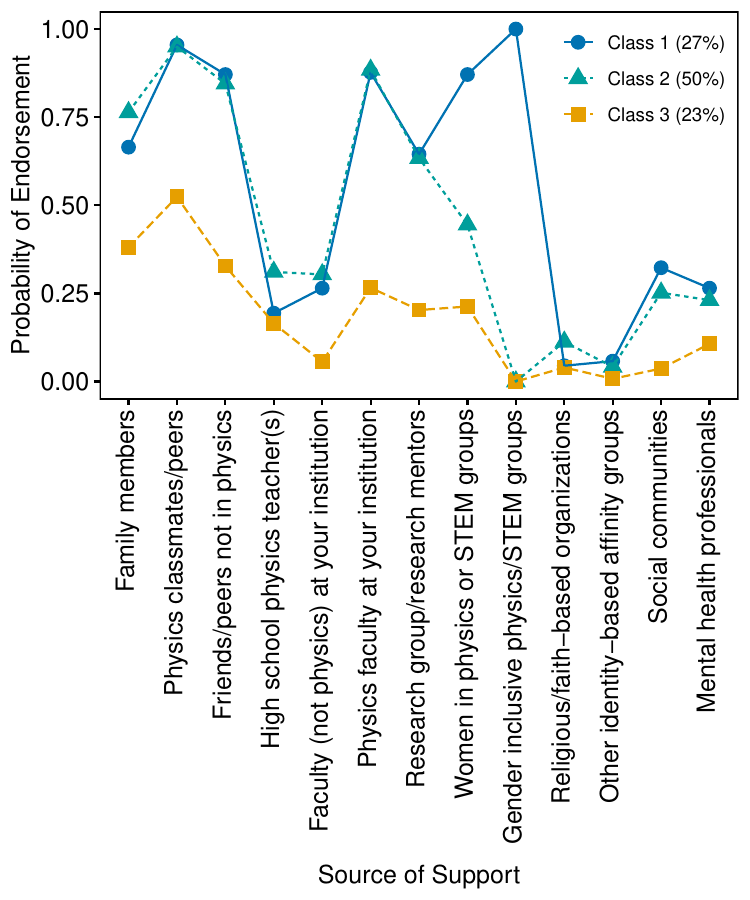}
    \caption{Three-class solution for latent class analysis. Percentages in the legend indicate the estimated percent of students (\textit{n} = 567) belonging to each class.}
    \label{fig:3Class}
\end{figure}

Based on the estimated item-endorsement probabilities (Fig.~\ref{fig:3Class}), we label the three classes as follows: Class 1---“High professional and identity-based support” (\textit{n} = 155, 27.3\%, 95\% C.I. [23.6\%, 30.9\%]), Class 2---“High professional and low identity-based support” (\textit{n} = 282, 49.7\%, 95\% C.I. [41.4\%, 56.1\%]), and Class 3---“Low professional and identity-based support” (\textit{n} = 130, 22.9\%, 95\% C.I. [16.8\%, 30.5\%]).

Students in the High professional and identity-based support class have a high probability of reporting diverse sources of support, including gender inclusive physics/STEM groups (100.0\%), physics classmates/peers (95.5\%), physics faculty (87.7\%), friends outside of physics (87.1\%), and family members (66.5\%). Students in this class, therefore, are well integrated into institutional, personal, and gender identity-based support networks during their undergraduate physics program.

Students in the High professional and low identity-based support class demonstrate a similar combination of social support to the High professional and identity-based support class (e.g., 95.0\% probability of receiving support from physics classmates/peers, 88.4\% probability of receiving support from physics faculty, and 76.4\% probability of receiving support from family members). However, these students are much less likely to receive support from women in physics or STEM groups (44.5\%) and receive virtually no support from gender-inclusive physics/STEM groups. 

Students in the Low professional and identity-based support class are characterized by consistently low probabilities of receiving support across all sources. Students in this class show only low-to-moderate likelihood of receiving support from personal networks (e.g., 38.1\% probability of receiving support from family members and 32.8\% probability of receiving support from non-physics peers), institutional mentorship (e.g., 26.6\% probability of receiving support from physics faculty), and identity-based groups (e.g., 21.3\% probability of receiving support from women in physics or STEM groups and virtually no support from gender-inclusive physics or STEM groups).


As mentioned earlier, LCA benefits from several posterior model classification diagnostics to assess the model fit. To calculate these diagnostics (Table~\ref{diagnostics}), we assigned each student to the class with the highest posterior probability. These class assignments are not definitive or absolute, nor carried forward in our subsequent analysis (in the next subsection); rather, they allow us to evaluate how well the model captures the structure of the data.
First, we calculated entropy, a measure of overall precision or reliability of classification for the whole sample across all the identified classes. The three-class model has an entropy of 0.87 (values range from zero to one, with higher values indicating better precision), indicating that the model has a reliable classification precision~\cite{masyn2013latent-key}. We also calculated the modal class assignment proportion (mcaP), the proportion of individuals in the sample who are assigned to each class based on their highest posterior probability. For the three-class model, the mcaP values all fell within the 95\% confidence intervals of the class proportions (which are “softer” estimates of the class sizes based on posterior probabilities), indicating minimal discrepancies and low class assignment errors. We also determined the average posterior class probabilities (AvePP) for individuals assigned to each of the three classes. As all AvePP values were greater than 0.8 (1.00, 0.94 and 0.88, respectively), the classification is considered good~\cite{masyn2013latent-key}. Finally, we calculated the odds of correct classification (OCC) for each class, which essentially compares our class assignment to random assignment. All three OCC values were greater than 5, indicating adequate class separation (i.e., the classes are distinct from one another) and classification precision~\cite{nagin2005group}. Taken together, these metrics suggest that the three-class model demonstrates good classification accuracy.

\begin{table}[t]
  \caption{Model classification diagnostics for the three-class solution of the latent class analysis. Brackets indicate 95\% confidence intervals.\label{diagnostics}}
  \begin{ruledtabular}
    \begin{tabular}{lcccc}
\textit{k}-class & \textit{k}-class proportions  & mcaP & AvePP & OCC\\ 
 \hline
 Class 1 &  0.273 [0.236, 0.309] & 0.273 & 1.000 & Inf \\ 
  Class 2 &  0.497 [0.414, 0.561] & 0.497 & 0.944 & 17.06\\
   Class 3 & 0.229 [0.168, 0.305] & 0.229 & 0.879 & 24.46\\
    \end{tabular}
  \end{ruledtabular}
   \begin{tablenotes}
      \small
      \item Note: mcaP = modal class assignment proportion; AvePP = average posterior class probabilities; OCC = odds of correct classification.
    \end{tablenotes}
\end{table}

\subsubsection{Relating social support class membership to gender identity and physics identity}

\begin{figure}[t]
    \centering
    \includegraphics[width=3.4in,trim={0 7cm 15cm 0 }]{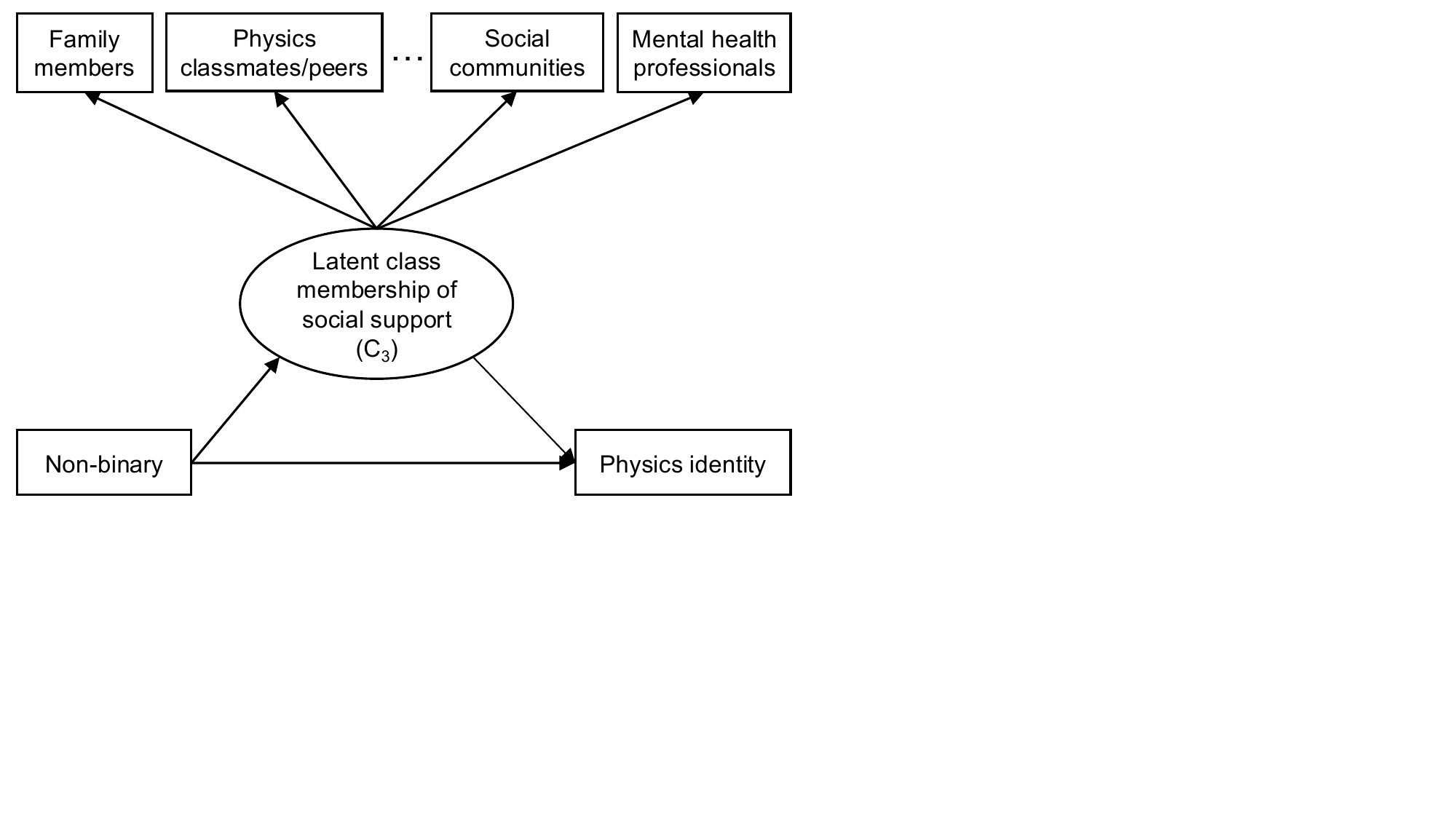}
    \caption{Path diagram of the model we used to measure the relationship between latent class membership (for the three-class solution, $C_3$; see Fig.~\ref{fig:3Class}) and two auxiliary variables: non-binary status and physics identity.}
    \label{fig:Path}
\end{figure}

Another advantage of LCA (and mixture modeling more broadly) is that it allows for integrating the identified classes into a larger system of auxiliary variables to understand how the emergent classes relate to other measured variables. Here we demonstrate an example of this to address the second research question. We include one of the gender identities (non-binary) as a covariate and physics identity as a distal outcome to understand how students’ latent class membership is related to these other variables (Fig.~\ref{fig:Path}). 

To estimate the model, we used the ML 3-step method~\cite{vermunt2010latent}, which allows covariates and distal outcomes to be simultaneously estimated while accounting for classification errors. In this model, we simultaneously estimated the effect of non-binary status on latent class membership and the effect of latent class membership on physics identity, while controlling for the direct relationship between non-binary status and physics identity. We estimated the likelihood of a student being in one class relative to a reference class based on a one-unit change in the covariate, non-binary status. We then estimated the conditional means for each latent class to evaluate the relationship between the latent class membership and the outcome variable, physics identity, controlling for non-binary status. The overall association was first evaluated using an omnibus Wald test and then using pairwise comparisons across the classes to explore class-specific differences.

Results indicate that students who identify as non-binary are more likely to be in Class 1, the High professional and identity-based support class, compared to Class 2, the High professional and low identity-based support class (logit = .99, $p <$ 0.001, Odds Ratio [OR] = 2.71). However, there is no significant difference in the probability of non-binary students being in Class 3, the Low professional and identity-based support class, compared to Class 1 (logit = --0.54, $p$ = 0.10, OR = 0.59) or in the probability of non-binary students being in Class 3 compared to Class 2 (logit = 0.46, $p$ = 0.21, OR = 1.59).

An omnibus Wald test indicates a significant difference in physics identity across the identified latent classes ($\chi^2$(2) = 6.09, $p <$ 0.05). Pairwise comparisons indicate that, after adjusting for the effect of non-binary status on physics identity, students in Class 1, the High professional and identity-based support class (\textit{M} = 2.60), and Class 2, the High professional and low identity-based support class (\textit{M} = 2.60), have the highest physics identity, followed by Class 3, the Low professional and identity-based support class (\textit{M} = 2.34). The difference in physics identity between Class 1 and Class 2, however, is not statistically significant ($M_{\text{diff}}$ = 0.004,  $p$ = 0.97). The differences in physics identity between Class 1 and Class 3 ($M_{\text{diff}}$ = 0.262,  $p$ = 0.02) and between Class 2 and Class 3 ($M_{\text{diff}}$ = 0.258,  $p$ = 0.02) are significant.

\begin{table*}[t]
  \caption{Pairwise comparison of subgroup assignment from \textit{k}-modes clustering (two-cluster solution; Fig.~\ref{fig:2Cluster}) and LCA (three-class solution; Fig.~\ref{fig:3Class}). Percentages indicate the percent of students in each LCA class that are assigned to each \textit{k}-modes cluster (i.e., columns add to 100\%; \textit{n} = 567).\label{comparison}}
  \begin{ruledtabular}
    \begin{tabular}{lcccc}
 & LCA Class 1:  & LCA Class 2: & LCA Class 3: & \\ 
  &  High professional and 
  & High professional and low 
 & Low professional and &\\ 
 & identity-based support & identity-based support  & identity-based support & Total\\
 \hline
\textit{K}-modes Cluster 1: High professional and &  149 (96.1\%) &  84 (29.8\%) &  8 (6.2\%) & 241\\ 
\hspace{2.5cm} identity-based support &&&& \\
\textit{K}-modes Cluster 2: High professional and low  &  6 (3.9\%) &  198 (70.2\%) &  122 (93.8\%)& 326 \\  
\hspace{2.5cm} identity-based support &&&& \\
 \hline
 Total & 155 & 282 & 130 & 567\\
    \end{tabular}
  \end{ruledtabular}
\end{table*}

\subsection{Direct comparison of \emph{k}-modes clusters and LCA classes
}


Using the modal class assignments from the LCA mentioned above (i.e., based on posterior probability), we compared how students’ class/cluster membership varied using each method. LCA and \emph{k}-modes clustering have a moderate level of agreement in our study (Table~\ref{comparison}). The first LCA class, the High professional and identity-based support class, aligns strongly with \emph{k}-modes Cluster 1 (also labeled as High professional and identity-based support), as 96\% of students in Class 1 were also assigned to Cluster 1. The second LCA class, the High professional and low identity-based support class, has moderate alignment with \emph{k}-modes Cluster 2 (also labeled as High professional and low identity-based support), with 70\% of students in this class being assigned to Cluster 2 and 30\% of students in this class being assigned to Cluster 1 (High professional and identity-based support). The third LCA class, Low professional and identity-based support class, is aligned well with \emph{k}-modes Cluster 2 (High professional and low identity-based support), with 94\% of students in this class being assigned to Cluster 2. Both methods, therefore, distinguish students who have access to diverse sources of support from those who report more limited sources of support. The three-class LCA model creates an additional partition among students with low identity-based support, creating one class of such students with high professional support (Class 2) and one class of such students with low professional support (Class 3).


It is important to acknowledge that the solutions we arrived at in this comparison are not definitive. We used the current best practices for each clustering method to carry out the analyses separately, as a researcher would have if they selected to use one of the methods.  For \emph{k}-modes clustering, we determined the optimal number of clusters based on the elbow plots of total within-cluster similarity and average silhouette width. We identified the optimal solution for the LCA models using multiple model fit indices, including BIC, aBIC, and VLMR-LRT.  To account for potential alternative model specifications and to offer transparency for readers interested in model comparison, we have included additional comparisons of the two methods in the Appendix (Figs.~\ref{fig:3Cluster} and~\ref{fig:2Class} and Tables~\ref{comparisonboth3} and~\ref{comparisonboth2}). These include results for the two-class solution in LCA and the three-cluster solution in \emph{k}-modes clustering and direct comparisons of student assignment to classes and clusters (as in Table~\ref{comparison}) when we impose the same number of classes/clusters between the two methods (i.e., comparing the two-class LCA solution to the two-cluster \emph{k}-modes solution and the three-class LCA solution to the three-cluster \emph{k}-modes solution).

\section{Summary and Conclusion}

Our theoretical descriptions and worked example indicate that both \emph{k}-modes clustering and LCA are useful when identifying heterogeneity within a population. LCA, however, offers the important advantage of directly incorporating auxiliary variables within the broader mixture modeling framework. In contrast to \emph{k}-modes clustering, LCA takes into account classification error, which reduces measurement bias and improves the validity of inferences. That is, when the LCA algorithm estimates that an individual has, for example, a 0.5 probability of being in Class 1, a 0.3 probability of being in Class 2, and a 0.2 probability of being in Class 3, it does not assign this individual to Class 1. Instead, this probabilistic information is retained in the following analysis. Depending on the analytic approach used (e.g., the ML three-step)~\cite{vermunt2010latent}, the model can estimate the parameters between the latent class variable and the auxiliary variables while properly adjusting for the fact that class assignment is not absolute.

In our worked example, such an analysis indicated that students identifying as non-binary were significantly more likely to belong to the High professional and identity-based support class compared to the High professional and low identity-based support class. Interestingly, there were no significant differences in students’ sense of physics identity between these two classes. We again note that we did not aim to completely address the research questions in the worked example. We recommend for future research to examine the relationship between combinations of social support and all gender identities (we only investigated non-binary status here for demonstration purposes) and other demographic variables (e.g., racial/ethnic identities) as well as other outcomes that may be impacted by students' social support networks (e.g., performance, retention in the physics major, and sense of belonging) to make more robust conclusions.

This study highlights the potential of LCA (and other mixture models) in physics education research to understand meaningful population heterogeneity and to examine how such heterogeneity relates to other variables of interest. We hope that our step-by-step example offers practical guidance for researchers interested in using LCA to generate rigorous findings in PER. In addition to LCA, other mixture modeling techniques such as latent profile analysis, latent transition analysis, and factor mixture modeling are powerful extensions for modeling continuous indicators, examining developmental changes over time, and integrating latent structures with measurement models, respectively~\cite{spurk2020latent,nylund2023ten,lubke2005investigating}. These mixture modeling tools can enable researchers to adopt a person-centered approach, one that acknowledges variation in students’ educational experiences.

\section*{ACKNOWLEDGEMENTS}

We thank all the survey participants and conference organizers including faculty, students, and American Physical Society staff. This material is based upon work supported by the National Science Foundation under Grant No. 2224786. M. S. is partly funded by the Cotswold Foundation Postdoctoral Fellowship at Drexel University.

\bibliography{Support.bib} 

\clearpage

\section*{APPENDIX}

\begin{table*}[t]
  \caption{Model convergence and Log-Likelihood (LL) replication rates using 200 initial and 100 final random starts for one- to seven-class models in the latent class analysis. Percentages for final starting value sets converging are out of the number of final random starts (in this case, 100), for LL replication are out of the number of final starting value sets converging, and for smallest class are out of the number of observations (\textit{n} = 567). ``Npar" indicates the number of parameters in the model. \label{200con}}
  \begin{ruledtabular}
    \begin{tabular}{lcccccccc}
   & &  & \multicolumn{2}{c}{Final starting value sets converging} & \multicolumn{2}{c}{LL replication} & \multicolumn{2}{c}{Smallest class}\\
   \cline{4-5}
\cline{6-7} \cline{8-9}  
 Model & Best LL  & npar & \textit{n} & \%  & \textit{n} & \% & \textit{n} & \% \\ 
 \hline
1-Class & --3,840 & 13 & 100 & 100\% & 100 & 100.0\% & 567 & 100\%\\
2-Class & --3,648 &  27 & 98 & 98\% & 96 & 98.0\% & 148 & 26.1\%\\
3-Class & --3,597 & 41 & 85 & 85\% & 81 & 95.3\% & 130 & 22.9\% \\
4-Class & --3,577 & 55 & 40 & 40\% & 18 & 45.0\% & 101 & 17.8\% \\
5-Class & --3,559 & 69 & 37 & 37\% & 1 & 2.7\% & 20 & 3.5\%\\
6-Class & --3,542 & 83 & 31 & 31\% & 4 & 12.9\% & 18 & 3.2\%\\
7-Class & --3,529 & 97 & 24 & 24\% & 1 & 4.2\% & 10 & 1.8\%\\
    \end{tabular}
  \end{ruledtabular}
\end{table*}

\subsection{Model convergence rates for latent class analysis}

Tables~\ref{200con} and~\ref{2000con} show the convergence and log-likelihood replication rates for different numbers of initial and final random starts of the models. As mentioned in the main text, the log-likelihood replication rates increase for the four- to seven-class models when 2000 initial and 500 final random starts are used instead of 200 initial and 100 final random starts.

\begin{table*}[t]
  \caption{Model convergence and Log-Likelihood (LL) replication rates using 2000 initial and 500 final random starts for four- to seven-class models in the latent class analysis. Percentages for final starting value sets converging are out of the number of final random starts (in this case, 500), for LL replication are out of the number of final starting value sets converging, and for smallest class are out of the number of observations (\textit{n} = 567). ``Npar" indicates the number of parameters in the model.  \label{2000con}}
  \begin{ruledtabular}
    \begin{tabular}{lcccccccc}
  & &  & \multicolumn{2}{c}{Final starting value sets converging} & \multicolumn{2}{c}{LL replication} & \multicolumn{2}{c}{Smallest class}\\
   \cline{4-5}
\cline{6-7} \cline{8-9}  
 Model & Best LL  & npar & \textit{n} & \% & \textit{n} & \% & \textit{n} & \% \\ 
 \hline
4-Class & --3,577 & 55 & 113 & 23\% & 39 & 34.5\% & 101 & 17.8\%\\
5-Class & --3,559 & 69 & 87 & 17\% & 9 & 10.3\% & 20 & 3.5\%\\
6-Class & --3,542 & 83 & 53 & 11\% & 10 & 18.9\% & 18 & 3.2\%\\
7-Class & --3,527 & 97 & 41 & 8\% & 1 & 2.4\% & 17 & 3.0\%\\
    \end{tabular}
  \end{ruledtabular}
\end{table*}

\subsection{Information criteria plot for latent class analysis}

\begin{figure}[b]
    \centering
    \includegraphics[width=3.4in,trim={0 0 0 0 }]{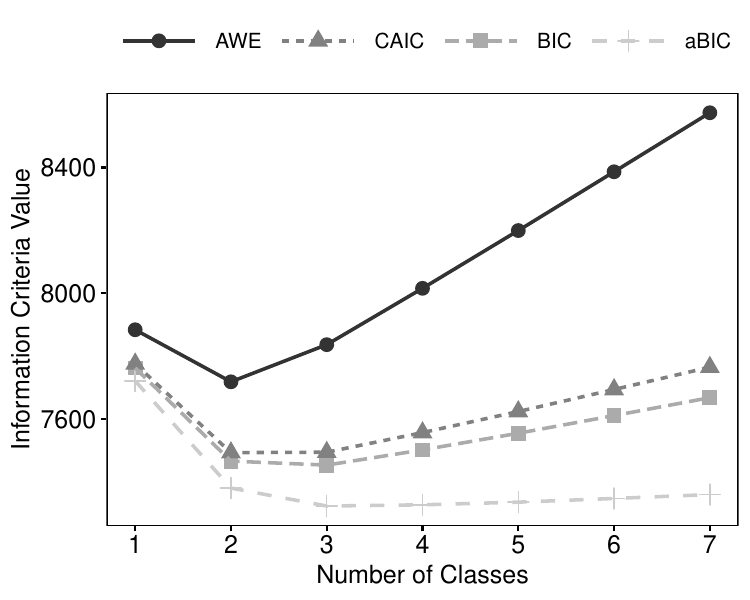}
    \caption{Information criteria plot for latent class analysis (values in Table~\ref{modelfit}).}
    \label{fig:IC}
\end{figure}

Figure~\ref{fig:IC} shows the information criteria values for one- to seven-class models for the latent class analysis. As mentioned in the main text, lower values are superior. The elbow point of the plot, where increasing the number of classes  does not significantly reduce the information criteria value, suggests that the three-class solution is optimal.

\subsection{Three-cluster solution for \textit{k}-modes clustering}

\begin{figure}[b]
    \centering
    \includegraphics[width=3.4in,trim={0 0 0 0 }]{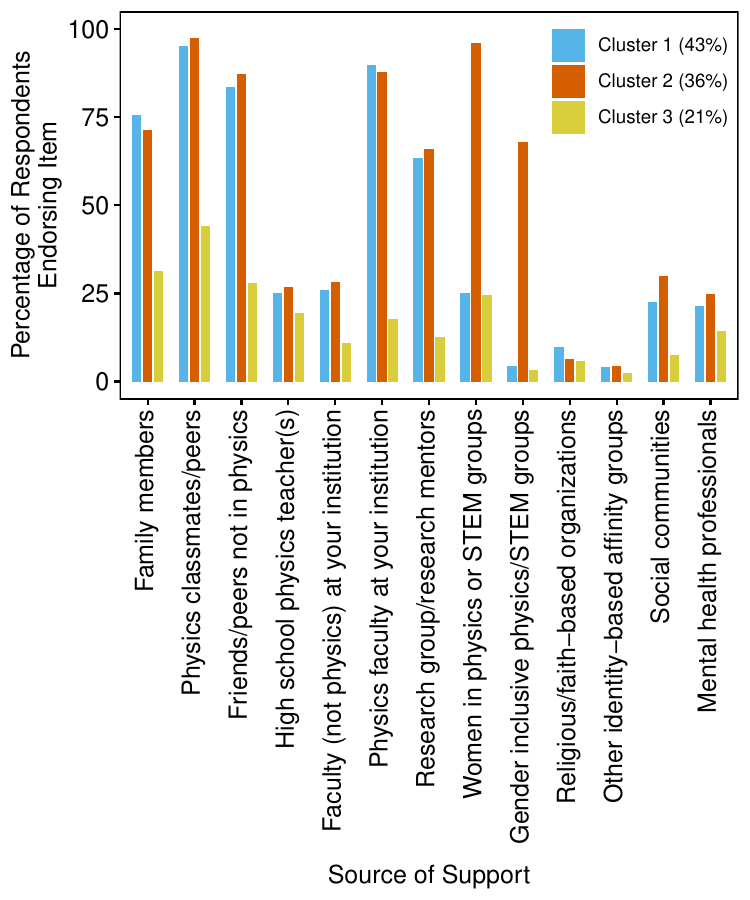}
    \caption{Three-cluster solution for \emph{k}-modes clustering. Percentages in the legend indicate the percent of students assigned to each cluster (\textit{n} = 567).}
    \label{fig:3Cluster}
\end{figure}

Figure~\ref{fig:3Cluster} shows the three-cluster solution for the \textit{k}-modes clustering method. Cluster 1 contains 43\% of the sample (\textit{n} = 243) and we label this cluster as ``High professional and low identity-based support." Similar to Cluster 2 in the two-cluster solution presented in the main text, many students in this cluster report receiving social support from professional sources (e.g., physics faculty at their institution). A low percentage of students in this cluster, however, report receiving support from women in physics or STEM groups and other gender inclusive physics/STEM groups.

Cluster 2 contains 36\% of the sample (\textit{n} = 206) and we label this cluster as ``High professional and identity-based support." Similar to Cluster 1 in the two-cluster solution presented in the main text, many students in this cluster report receiving social support from both professional (e.g., physics faculty at their institution) and identity-based (e.g., women in physics or STEM groups) sources.

Cluster 3 contains 21\% of the sample (\textit{n} = 118) and we label this cluster as ``Low professional and identity-based support." Most students in this cluster do not report receiving support from either professional or identity-based sources.

\subsection{Two-class solution for latent class analysis}

\begin{figure}[t]
    \centering
    \includegraphics[width=3.4in,trim={0 0 0 0 }]{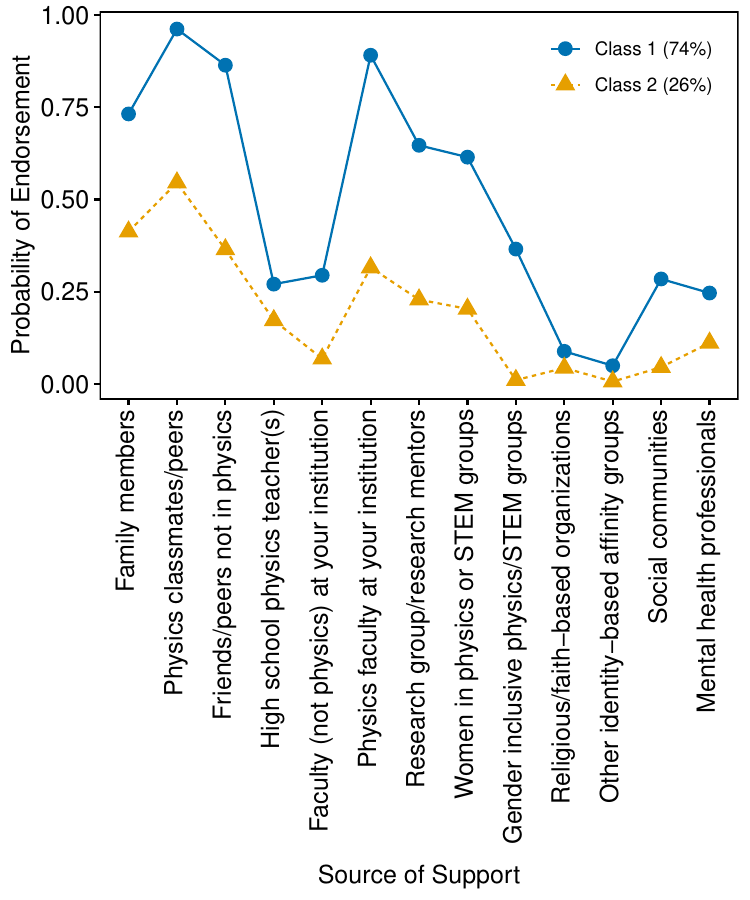}
    \caption{Two-class solution for latent class analysis. Percentages in the legend indicate the estimated percent of students belonging to each class (\textit{n} = 567).}
    \label{fig:2Class}
\end{figure}

Figure~\ref{fig:2Class} shows the two-class solution using LCA. Class 1 contains 74\% of the sample (\textit{n} = 427) and we label this cluster as ``High professional and identity-based support." Students in this cluster have a high probability of reporting support from both professional and identity-based sources.

Class 2 contains 26\% of the sample (\textit{n} = 140) and we label this cluster as ``Low professional and identity-based support." Students in this cluster have a low probability of reporting support from both professional and identity-based sources.

\subsection{Pairwise comparison of subgroup assignment for two-cluster vs. two-class solutions}

Table~\ref{comparisonboth2} compares students' subgroup assignment when two-cluster/class solutions are imposed for both methods. Despite the similar labels and characteristics between Class 1 and Cluster 1 (High professional and identity-based support), only about half of the students in Class 1 belong to Cluster 1. The other half of Class 1 is assigned to Cluster 2 (High professional and low identity-based support). There is strong alignment between Class 2 (Low professional and identity-based support) and Cluster 2, with about 94\% of students in Class 2 being assigned to Cluster 2. Therefore, the two algorithms produce slightly different classifications when creating two subgroups. \textit{K}-modes clustering mainly separates students with high versus low identity-based support (i.e., students in both clusters report high professional support), while LCA disentangles students based on differences in both professional and identity-based support.

\begin{table*}[t]
  \caption{Pairwise comparison of subgroup assignment from \textit{k}-modes clustering and LCA when a two-cluster/class solution is imposed for both methods (Figs.~\ref{fig:2Cluster} and~\ref{fig:2Class}). Percentages indicate the percent of students in each LCA class that are assigned to each \textit{k}-modes cluster (i.e., columns add to 100\%; \textit{n} = 567).\label{comparisonboth2}}
  \begin{ruledtabular}
    \begin{tabular}{lccc}
 & LCA Class 1:  & LCA Class 2: & \\ 
 & High professional and & Low professional and  & \\
  & identity-based support & identity-based support  & Total\\
 \hline
\textit{K}-modes Cluster 1: High professional and & 232 (54.3\%) & 9 (6.4\%) & 241\\ 
\hspace{2.5cm} identity-based support &&&\\
\textit{K}-modes Cluster 2: High professional and low &  195 (45.7\%) & 131 (93.6\%) & 326\\ 
\hspace{2.5cm} identity-based support &&&\\
 \hline
 Total & 427 & 140 &567\\
    \end{tabular}
  \end{ruledtabular}
\end{table*}

\begin{table*}[t]
  \caption{Pairwise comparison of subgroup assignment from \textit{k}-modes clustering and LCA when a three-cluster/class solution is imposed for both methods (Figs.~\ref{fig:3Class} and~\ref{fig:3Cluster}). Percentages indicate the percent of students in each LCA class that are assigned to each \textit{k}-modes cluster (i.e., columns add to 100\%; \textit{n} = 567).\label{comparisonboth3}}
  \begin{ruledtabular}
    \begin{tabular}{lcccc}
 & LCA Class 1:  & LCA Class 2: & LCA Class 3: & \\ 
  &  High professional and 
  & High professional and low 
 & Low professional and &\\ 
 & identity-based support & identity-based support  & identity-based support & Total\\
 \hline
\textit{K}-modes Cluster 1: High professional and low &  11 (7.1\%) &  209 (74.1\%) &  23 (17.7\%) & 243\\ 
\hspace{2.5cm} identity-based support &&&&\\
\textit{K}-modes Cluster 2: High professional and &  140 (90.3\%) &  66 (23.4\%) &  0 (0\%)& 206 \\ 
\hspace{2.5cm} identity-based support &&&&\\
\textit{K}-modes Cluster 3: Low professional and &  4 (2.6\%) &  7 (2.5\%) &  107 (82.3\%)& 118 \\ 
\hspace{2.5cm} identity-based support &&&&\\
 \hline
 Total & 155 & 282 & 130 & 567\\
    \end{tabular}
  \end{ruledtabular}
\end{table*}

\subsection{Pairwise comparison of subgroup assignment for three-cluster vs. three-class solutions}

Table~\ref{comparisonboth3} compares students' subgroup assignment when three-cluster/class solutions are imposed for both methods. There is strong alignment between the two methods when both algorithms are prompted to identify three subgroups. Indeed, we give the three subgroups the same labels in both cases. Class 1 and Cluster 2 both indicate students with high professional and identity-based support, and 90\% of students in Class 1 are assigned to Cluster 2. Class 2 and Cluster 1 both indicate students with high professional and low identity-based support, and 74\% of students in Class 2 are assigned to Cluster 1. Finally, Class 3 and Cluster 3 both indicate students with low professional and identity-based support, and 82\% of students in Class 3 are assigned to Cluster 3. 

\end{document}